# Ferroelectric-tunable quantum nonlinearity of chiral Bloch electrons in a moiré system

Zitian Pan[1,2†], Jundong Zhu[1,2†], Yu Hong[1,2], Jingwei Dong[1,2], Dongxia Shi[1,2], Kenji Watanabe[3], Takashi Taniguchi[4], Luojun Du[1,2], Wei Yang[1,2*], Guangyu Zhang[1,2*]

[1]*Beijing National Laboratory for Condensed Matter Physics and Institute of Physics, Chinese Academy of Sciences, Beijing 100190, China*
[2]*School of Physical Sciences, University of Chinese Academy of Sciences, Beijing 100190, China*
[3]*Research Center for Functional Materials, National Institute for Materials Science, 1-1 Namiki, Tsukuba 305-0044, Japan*
[4]*International Center for Materials Nanoarchitectonics, National Institute for Materials Science, 1-1 Namiki, Tsukuba 305-0044, Japan*
[†]*These authors contributed equally to this work.*
**Corresponding author: wei.yang@iphy.ac.cn; gyzhang@iphy.ac.cn*

**Abstract:** Sliding ferroelectricity in van der Waals materials shows great potential for designing robust memory devices. However, its thermodynamic behaviors and the coupling with certain quantum effects remain largely unexplored. Here, we demonstrate ferroelectric control over quantum nonlinear transport in a hexagonal boron nitride (hBN) encapsulated twisted double-bilayer graphene moiré heterostructure. The ferroelectricity is attributed to the presence of rhombohedral stacking in the top hBN, confirmed by both electrical transport and optical second harmonic generation (SHG) measurements. Remarkably, the polarization magnitude remains temperature-independent across 1.7-200 K, while nucleation time exhibits thermally activated behavior, decreasing with increasing temperature. Furthermore, we demonstrate a ferroelectric-switchable nonlinear Hall effect, attributed to the chiral scattering induced by Berry curvature, with outstanding fatigue-resistant and nonvolatility, demonstrating direct coupling between sliding ferroelectricity and quantum geometric properties. Our results establish sliding ferroelectrics as a platform for exploring electrically programmable Berry curvature physics.

Ferroelectricity, the ability to reverse spontaneous polarization with an external field, plays a critical role in modern electronics, showing great potential in non-volatile memories, sensors, and actuators[1-3]. In conventional ferroelectrics, switching proceeds via domain nucleation and wall motion, and its kinetics have been extensively studied over decades[4-9]. The recent emergence of sliding ferroelectricity in van der Waals materials offers a paradigm fundamentally different from conventional displacive ferroelectrics: polarization switching arises from interlayer sliding rather than ionic displacement, potentially decoupling the switching dynamics from the lattice[10-21]. However, despite rapid progress, the realization of robust sliding ferroelectricity coupled with distinctive quantum transport effects and its fundamental thermodynamic behaviors, particularly the nucleation processes that govern switching speed and reliability, remain largely unexplored in this new class of ferroelectrics.

Parallel to these developments, moiré systems formed by stacking two-dimensional materials with a slight twist or lattice mismatch have emerged as a powerful platform for engineering quantum nonlinearities[22-30]. In such systems, the long-wavelength superlattice potential reconstructs the electronic band structure, giving rise to flat bands, strong correlations, and nontrivial Berry curvature[31-39]. A hallmark of these engineered bands is the nonlinear Hall effect, which can go beyond the constraint of linear charge transport and offers a unique window into the quantum geometry of Bloch electrons. For example, it provides a direct probe for Berry curvature dipole/multipole, Berry connection polarizability, and quantum metric dipole[27,30,40-51]. Among various moiré platforms, twisted double-bilayer graphene (TDBG) stands out for its exceptional tunability and rich quantum phenomena, making it an ideal platform for studying the coupling of ferroelectricity and quantum effects[28-30,36-38,52-55].

Here, we bridge these two frontiers, sliding ferroelectricity and moiré quantum nonlinearity, by demonstrating ferroelectric control over the nonlinear Hall response in a hBN/TDBG/hBN heterostructure. Through systematic electrical transport and optical second harmonic generation (SHG) measurement, we attribute the ferroelectricity to the presence of rhombohedral stacking in the top hBN layer. Notably, the polarization magnitude remains constant across 1.7–200 K, while the nucleation time displays thermally activated behavior, decreasing with increasing temperature. To our best knowledge, direct measurement of the nucleation time in sliding ferroelectrics has not been reported. The decoupling between static polarization stability and dynamic switching parameters points to an interfacial locking mechanism unique to sliding ferroelectrics. Furthermore, we observe a ferroelectric-tunable nonlinear Hall effect arising from the spontaneous inversion symmetry breaking in TDBG. The nonlinear Hall response remains bipolar switching after more than 1300 cycles and retention exceeding 7 hours at 1.7 K, demonstrating non-volatile control over Berry curvature. By harnessing the interplay between sliding ferroelectricity and moiré quantum geometry, our results establish a new pathway for electrically programmable Berry curvature engineering and open the door to devices where memory and quantum transport are intimately coupled.

## Ferroelectricity in High-Quality hBN/TDBG/hBN Device

We fabricate a high-quality dual-gate ABBA-TDBG device, in which two pieces of bilayer graphene are sandwiched by hBN, using $3/30$ nm Ti/Au and a few layers graphene for the top and the bottom gates, respectively (Fig. 1a). Optical SHG measurement shows that signals of SHG emerge only at top layer hBN, while absent at bottom layer hBN (Fig. 1b), suggesting a broken inversion symmetry in the former[56].

The dual-gate structure enables independent control of carrier density $n$ and displacement field $D$: $n = (C_b V_b + C_t V_t)/e$, $D = (C_b V_b - C_t V_t)/2\varepsilon_0$, where $C_b$ ($C_t$) is the capacitance per area of bottom (top) layer BN, $V_b$ ($V_t$) is the bottom (top) gate voltage, $e$ is the elementary charge, $\varepsilon_0$ is vacuum permittivity. Fig. 1c shows the longitudinal resistance $R_{xx}$ as a function of $V_b$ and $V_t$. The charge neutral point (CNP, $n = 0$) is identified by the high resistance states, indicated by the red color in Fig. 1c. Fig. 1e displays one Landau fan diagram at $T = 1.7$ K, and the moiré superlattice can be inferred from the emergence of Landau levels (LL) generated at high carrier densities $n_s \sim 11.5 \times 10^{12} \text{cm}^{-2}$. We obtain a twist angle of $\theta = 2.23°$ according to $n_s = 4/A \approx 8\theta^2/(\sqrt{3}a^2)$, where $A$ is the area of a moiré unit cell and $a$ is the lattice constant of graphene. The twist angle extracted from Landau fan diagrams of different Hall bars are the same, suggesting high uniformity of the device (Supporting Information Fig. S2 and Table S1). The Landau fan diagrams, together with the displacement field tuning of the LLs in Supporting Information Fig. S3, are consistent with the transport properties in high quality TDBG devices with a large twist angle [36,37,52].

Beside the common feature in large angle twisted TDBG, we observe a sudden shift of the CNP resistance peak near $V_t = 5.7$ V in Fig. 1c. We then change the scanning direction, i.e. keep $n$ unchanged and then sweep $D$ from negative to positive ($V_b$ from negative to positive and $V_t$ from positive to negative). As shown in Fig. 1d, a shift corresponding to a constant $V_t = -5.7$ V, instead of $D$ (also see Supporting Information Fig. S4 for $R_{xx}$ maps as a function of $n$ and $D$), is observed, suggesting a ferroelectricity related to the top gate. Here, the shift appears on the other side because of the opposite sweeping direction of $V_t$, consistent with previous reports and the properties of ferroelectricity[11,14]. It's worth noting that the ferroelectric switching remains unchanged under magnetic fields, as shown in Supporting Information Fig. S5.

Further measurements are carried out to confirm the origin of the ferroelectricity. Fig. 1f (g) shows the transfer curves as a function of top (bottom) gate in both forward and backward sweep direction under fixed bottom (top) gate. For the top gate sweeping at $V_b = 0$ V, a clear hysteresis is observed, and when changing the scanning speed, the hysteresis shows no significant difference (Fig. 1f). By contrast, two transfer curves overlap when sweeping back gate at $V_t = 0$ V (Fig. 1g).

Taking into account the above experimental observations, we attribute the observed ferroelectricity to sliding ferroelectricity originating from the rhombohedral stacking in top hBN. First, pronounced hysteresis is observed only in the top gate (Fig. 1f), while absent in the bottom gate (Fig. 1g). This rules out the possibility that the ferroelectricity originates from the bottom hBN or from the TDBG itself. Moreover, the hysteresis window remains unchanged when we vary the top-gate sweeping rate (Fig. 1f), which effectively excludes extrinsic mechanism such as charge trapping or surface dipole effects[57-59]. These observations leave the top hBN as the only plausible origin. Furthermore, the abrupt resistance jumps upon exceeding a threshold gate voltage (Fig.

1c and 1d) are fully consistent with previous reports of sliding ferroelectricity in rhombohedral-stacked hBN[11,60] and TMDCs[14,18,21]. The emergence of SHG signals in the top hBN (Fig. 1b) also confirms the broken inversion symmetry, which is a necessary condition for ferroelectricity. While an odd-layer hexagonal-stacked hBN can also produce an SHG signal, given its horizontal mirror symmetry, such a stacking configuration does not give rise to an out-of-plane ferroelectric polarization. Therefore, we attribute the observed ferroelectricity to sliding ferroelectricity induced by rhombohedral stacking in the top hBN layer.

In addition, it's worth noting that recent studies have reported unconventional ferroelectricity or electronic ratchet effect in graphene encapsulated with the same hBN crystals[61-64]. These effects have been attributed to line defects or stacking faults[65,66], indicating that complex stacking orders can occasionally occur even in high-quality bulk crystals. This lends additional plausibility to the presence of a rare rhombohedral stacking domain in our top hBN layer.

**Thermodynamics of Ferroelectricity**

For the emerging sliding ferroelectricity to reach its full potential, a comprehensive understanding of its thermodynamic properties is essential. Here, we focus on the regime where a ferroelectric switching occurs near CNP, which could result in a high switching on/off ratio. For a fixed back gate $V_b = -5.65\ \mathrm{V}$ , as shown in Fig. 2a, we observe a sudden jump of two orders of magnitude in resistance over the transition. We further study the switching process by sending a voltage pulse on the top gate (Fig. 2b). When the pulse is short, the resistance shows no response to the pulse. However, for a sufficiently long pulse, it first remains unchanged (highlighted by orange, process I), and then experiences a sudden jump (process II). When $V_t$ returns to a low level, $R_{xx}$ gradually decreases, and finally remains stable at a high resistance state (process III). Process I corresponds to a nucleation time, and it characterizes the shortest time for a pulse with a given amplitude to switch the ferroelectric polarization. Process II reflects an ultrafast switching process, and the reported speed can be up to $10^3\mathrm{m/s}$[18,60,67]. Process III is due to the carrier relaxation.

The critical width $t_c$ for a pulse to switch the polarization is highly related to the high level of the top gate pulse $V_H$, since the nucleation time decreases with the increase of pulse amplitude (Fig. 2b). As shown in Fig. 2c and Supporting Information Fig. S6, $t_c$ decreases exponentially with the increase of $V_H$, and can be well fitted by the equation:

$$t_c = t_0 \exp\left(-\frac{a}{V_H - V_0}\right)$$

where $t_0$ and $V_0$ characterize the nucleation time and coercive voltage, respectively, and $a$ is a fitting parameter. As the temperature increases, $t_0$ decrease (Fig. 2d), which means that nucleation is easier to take place at higher temperatures.

The shortest operation pulse obtained here is in the millisecond range, consistent with some previous reports[10,60]. However, some other reports show that the switching can occur at the nanosecond level[18,60,67]. This discrepancy arises from fundamental differences in measurement protocols and the corresponding physical processes being probed. The ferroelectric switching process contains two parts: domain nucleation and movement of the domain wall. In reports demonstrating nanosecond-switching, the switching was achieved by applying a single voltage pulse significantly exceeding the coercive field. Under such high overdrive conditions, domain wall motion is dominant and enters the superlubric regime[67], enabling sub-microsecond switching. This approach measures the upper limit of switching speed under ideal conditions. In contrast, our method systematically determines the critical pulse width for each pulse amplitude near the coercive field. This protocol probes the thermally activated switching regime, where polarization reversal is governed by the competition between the applied field and local pinning potentials. The extracted parameter $t_0$ characterizes the waiting time for domain nucleation, which is exponentially dependent on the energy barrier height[7-9]. The observed decrease of $t_0$ with increasing temperature (Fig. 2d) confirms the thermal activation nature of this process. Our results thus provide critical information for device operation near the coercive field, which is essential for low-power devices, and complement previous studies focused on high-speed switching limits.

In contrast to the thermally activated switching process, the static polarization state is independent of the temperature. Supporting Information Fig. S7 shows the transfer curves of the top gate in forward and backward sweep directions at different temperatures. The magnitude of the polarization $P_{2D}$ can be extracted from the hysteresis as $P_{2D} = C_t \Delta V_{t0} d_t / 2$, where $C_t$ is the capacitance per area of top hBN, $\Delta V_{t0}$ is the magnitude of hysteresis and $d_t$ is the thickness of top hBN. $P_{2D}$ is estimated to be about $2.96 \times 10^{-12}$ C m$^{-1}$, and is nearly unchanged under different temperature, confirming the interlayer polarization origin of the ferroelectricity[11].

**Ferroelectric-Tunable Nonlinear Hall Effect**

Next, we turn to explore the ferroelectric tuning of the quantum geometric effects in TDBG. Due to the broken inversion symmetry, TDBG has a nonzero Berry curvature at the band edge[33,35,68], and it becomes a new system for studying the nonlinear Hall effect[28-30]. The quantum nonlinearity can arise from the chiral scatterings of Bloch electrons via skew scatterings and side jumps[24,69,70]. If considering the breaking of $C_3$ symmetry in the moiré superlattice by strain and other breaking terms, nonlinear Hall effects could also due to Berry curvature dipole in TDBG[30,42].

The switchable polarization in top layer hBN together with the broken inversion symmetry in TDBG enable ferroelectric-tunable nonlinear Hall effects. Fig. 3a shows two plots of the second-order nonlinear Hall response $V_{xy}^{2\omega}$ against the top gate voltage, where the blue and the red correspond to the two opposite gate voltage sweeping directions, revealing a prominent hysteresis. $V_{xy}^{2\omega}$ switches its sign across the CNP, and it reaches a max response of 20 μV at a given excitation of $I = 10$ nA. Then we keep the gates fixed, and investigate the excitation electric-field dependence of the nonlinear response. As shown in Fig. 3b, the observed quadratic dependence of $V_{xy}^{2\omega}$ on $V_{xx}^{\omega}$, together with the same magnitude yet opposite sign when the source/drain configuration and the detection probes exchanges simultaneously, confirm the intrinsic second-order nonlinear nature. By applying a series of top gate pulses, the nonlinear Hall effect can be effectively turned on and off, as shown in Fig. 3c. Beside the ferroelectric switching, $V_{xy}^{2\omega}$ shows some random jumps in the "on" state, which may due to ferroelectric domain dynamics and relaxation. The stronger sensitivity in $V_{xy}^{2\omega}$, compared to the small change in resistance, suggests that the nonlinear transport could be a powerful probe for studying the domain wall dynamics in a sliding ferroelectric.

Moreover, such a switchable nonlinear response can also be useful for designing robust memory devices. The switching properties remains unchanged after switching for more than 1300 times, showing a good fatigue resistance (Fig. 3d). The retention property is also tested by applying a pulse to set and reset the device and then investigate the time dependence of the nonlinear response (Fig. 3e). Second order nonlinear response as well as longitudinal resistance at "on" and "off" states are nearly constant and can be clearly distinguished after more than 7 hours, indicating a nonvolatility of the polarization.

It's worth noting that the ferroelectric switching of $V_{xy}^{2\omega}$ is attributed to the shift of carrier density instead of a change in symmetry. As shown in Fig. 3a, $V_{xy}^{2\omega}$ exhibits hysteresis, which means that a nonzero $V_{xy}^{2\omega}$ can be observed under both the "up" and "down" polarization states, and it is merely shifted in gate voltage. Furthermore, switching the polarization of rhombohedral hBN, which corresponds to a relative translation of the two hBN layers by one third of the lattice constant, does not change the overall symmetry of the system.

**Conclusion**

Our work establishes the rhombohedral-stacking-hBN/TDBG heterostructure as a model system for investigating the sliding ferroelectricity, where the rhombohedral stacking hBN serves as the polarization origin while TDBG moiré system provides an ultrasensitive probe for electrical readout and platform for studying quantum effects. The coexistence of temperature-independent polarization and thermal activated switching dynamics offers critical insights for designing robust and low-power memory devices. The observed interplay between sliding-induced polarization and nonlinear transport further suggests new possibilities for electrically programmable Berry curvature engineering in van der Waals heterostructures.

**Supporting Information**

Device fabrication, electrical measurement, high quality of the device, and extended data for thermodynamics of the ferroelectricity.

**Acknowledgements:**

We acknowledge support from the Natural Science Foundation of China (NSFC, Grant No. 62488201), the Strategic Priority Research Program of Chinese Academy of Sciences (No. XDB1710000), the National Key Research and Development Program (Grant Nos. 2024YFA1410400, 2021YFA1202900), Quantum Science and Technology-National Science and Technology Major Project (Grant No. 2025ZD0300500), the Beijing Natural Science Foundation (JQ25003). K.W. and T.T. acknowledge support from the JSPS KAKENHI (Grant Numbers 20H00354, 21H05233 and 23H02052) and World Premier International Research Center Initiative (WPI), MEXT, Japan.

**Author Contributions:**

W.Y. and G.Z. supervised this work; Z.P., W.Y. and L.D. conceived the project and designed the experiments; J.Z. fabricated the devices; Z.P. carried out the optical and electrical measurements; K.W. and T.T. provided high quality h-BN crystals; Z.P., J.Z. and W.Y. analyzed the data; Y.H., J.D., D.S. and L.D. helped with data analysis; Z.P., W.Y. and G.Z. co-wrote the manuscript. All authors discussed the results and commented on the paper.

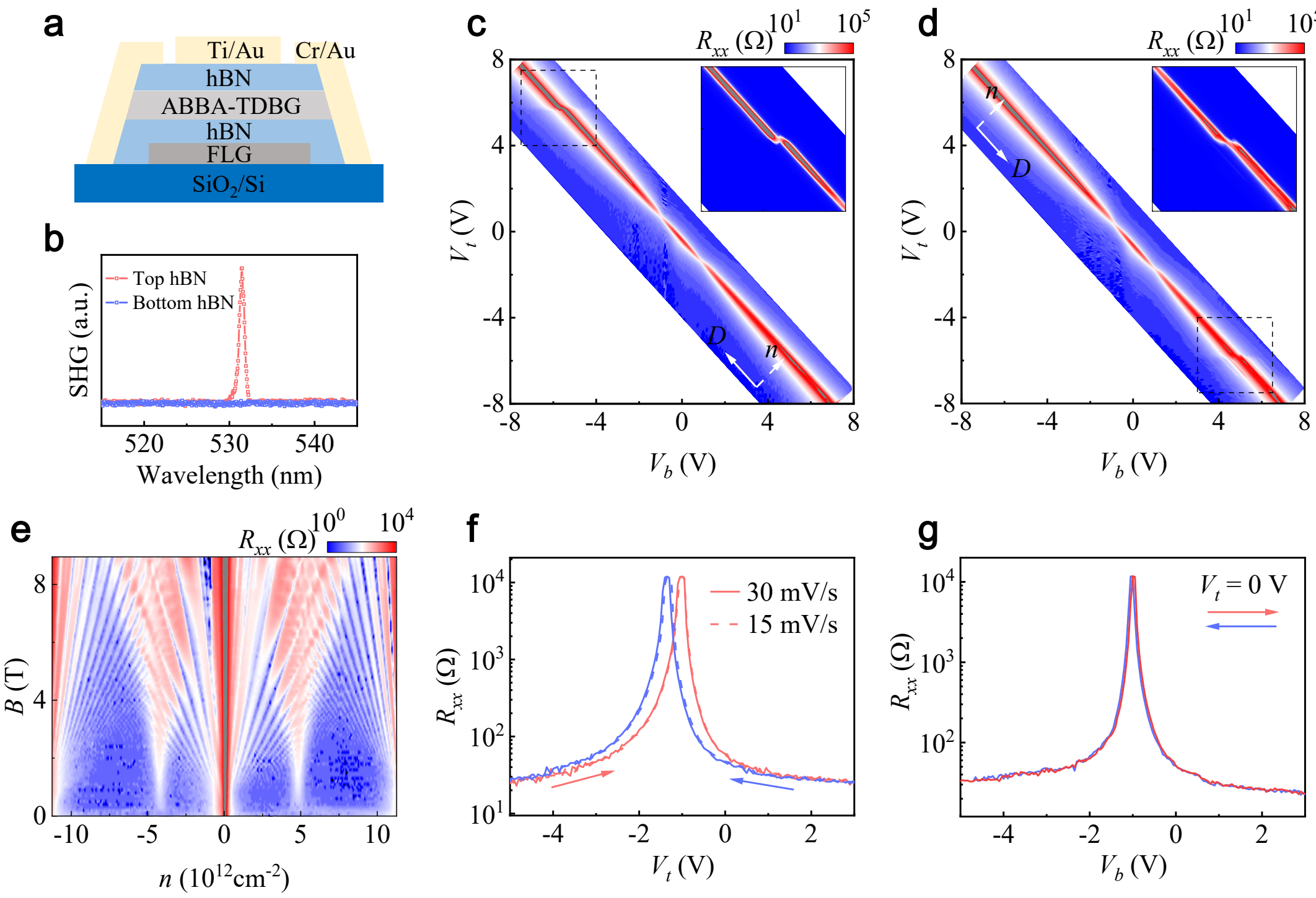

**Fig. 1. Ferroelectricity in high-equality hBN/TDBG device. a,** Schematic of the device structure. **b,** SHG response of top (red) and bottom (blue) layer hBN under excitation laser with wavelength of 1064 nm. **c,d,** Longitudinal resistance $R_{xx}$ as a function of top gate $V_t$ and bottom gate $V_b$. The arrows indicate the scanning direction, where the solid one is fast scan and the dashed one is slow scan. Inset: enlarged view of the dashed box region, plotted on linear scale. **e,** $R_{xx}$ as a function of carrier density $n$ and magnetic field $B$ under zero displacement field $D$. **f,** Transfer curves of $V_t$ in both forward and backward scanning direction under $V_b = 0$ V. **g,** Transfer curves of $V_b$ in both forward and backward scanning direction under $V_t = 0$ V.

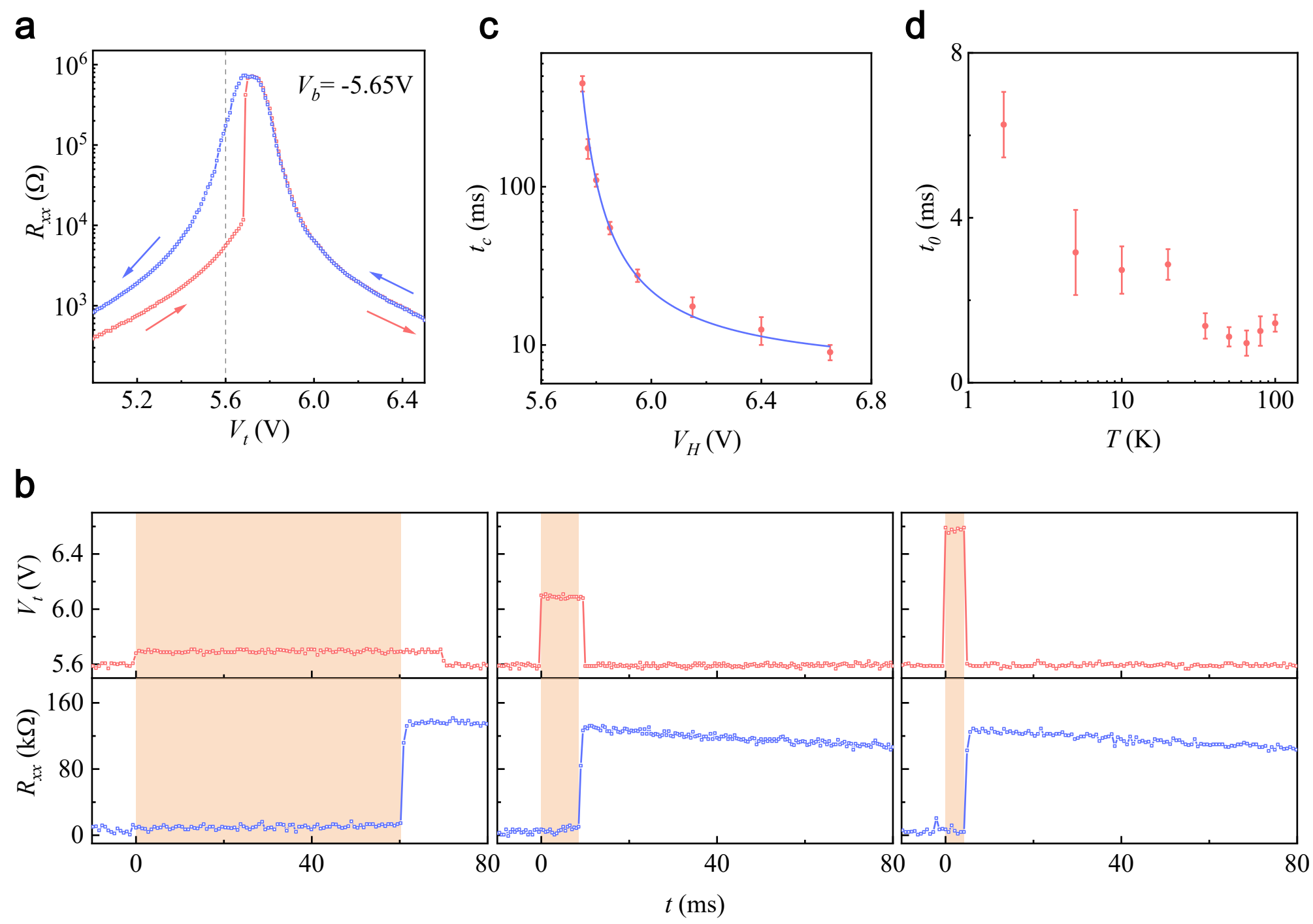

**Fig. 2. Thermodynamic behaviors of the ferroelectricity. a,** Transfer curves of $V_t$ in both forward and backward scanning direction under $V_b = -5.65$ V. **b,** Ferroelectric switching under pulses with different amplitude. **c,** Critical width $t_c$ to switch the polarization as a function of the high voltage level of the pulse $V_H$ at $T = 1.7$ K. The blue line is the fitting result with $t_c = t_0 \exp\left(-\frac{a}{V_H - V_0}\right)$. **d,** Temperature dependence of $t_0$.

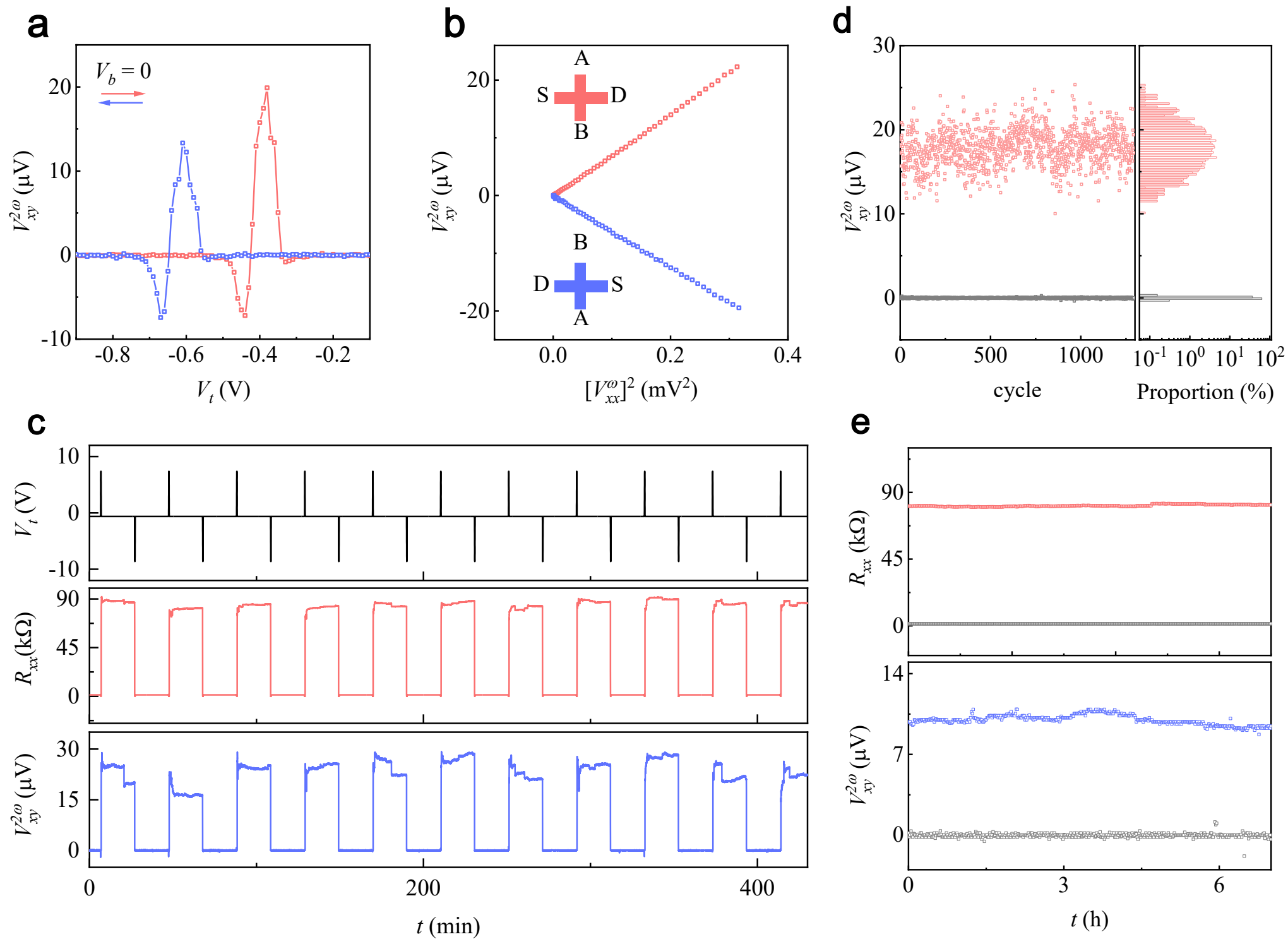

**Fig. 3. Ferroelectric-tunable nonlinear Hall effect. a,** Hysteresis of the nonlinear response $V_{xy}^{2\omega}$ when scanning $V_t$ in forward and backward directions. **b,** Excitation electric field dependence of $V_{xy}^{2\omega}$ in opposite source/drain and detection probes configuration. **c,** Ferroelectric switching of $R_{xx}$ and $V_{xy}^{2\omega}$ under a series of voltage pulses. **d,** Left: $V_{xy}^{2\omega}$ at "on" (red) and "off" (gray) states against switching cycles. Right: distribution of $V_{xy}^{2\omega}$. **e,** Retention property of the ferroelectric state.

## Reference

1 D. W. Zhang, P. Schoenherr, P. Sharma & J. Seidel. Ferroelectric order in van der Waals layered materials. *Nature Reviews Materials* **8**, 25-40 (2023).

2 P. Muralt. Ferroelectric thin films for micro-sensors and actuators: a review. *Journal of Micromechanics and Microengineering* **10**, 136-146 (2000).

3 M. Dawber, K. M. Rabe & J. F. Scott. Physics of thin-film ferroelectric oxides. *Reviews of Modern Physics* **77**, 1083-1130 (2005).

4 J. Y. Yang et al. Theoretical Lower Limit of Coercive Field in Ferroelectric Hafnia. *Physical Review X* **15**, 021042 (2025).

5 J. E. Huber & N. A. Fleck. Multi-axial electrical switching of a ferroelectric: theory versus experiment. *Journal of the Mechanics and Physics of Solids* **49**, 785-811 (2001).

6 V. Boddu, F. Endres & P. Steinmann. Molecular dynamics study of ferroelectric domain nucleation and domain switching dynamics. *Sci Rep* **7**, 806 (2017).

7 T. D. Cornelissen, I. Urbanaviciute & M. Kemerink. Microscopic model for switching kinetics in organic ferroelectrics following the Merz law. *Physical Review B* **101**, 214301 (2020).

8 T. D. Cornelissen et al. Kinetic Monte Carlo simulations of organic ferroelectrics. *Phys Chem Chem Phys* **21**, 1375-1383 (2019).

9 I. Urbanaviciute et al. Suppressing depolarization by tail substitution in an organic supramolecular ferroelectric. *Phys Chem Chem Phys* **21**, 2069-2079 (2019).

10 K. Ko et al. Operando electron microscopy investigation of polar domain dynamics in twisted van der Waals homobilayers. *Nature Materials* **22**, 992-998 (2023).

11 K. Yasuda, X. Wang, K. Watanabe, T. Taniguchi & P. Jarillo-Herrero. Stacking-engineered ferroelectricity in bilayer boron nitride. *Science* **372**, 1458-1462 (2021).

12 L. Molino et al. Ferroelectric Switching at Symmetry-Broken Interfaces by Local Control of Dislocations Networks. *Adv Mater* **35**, e2207816 (2023).

13 M. Vizner Stern et al. Interfacial ferroelectricity by van der Waals sliding. *Science* **372**, 1462-1466 (2021).

14 X. Wang et al. Interfacial ferroelectricity in rhombohedral-stacked bilayer transition metal dichalcogenides. *Nat Nanotechnol* **17**, 367-371 (2022).

15 A. Weston et al. Interfacial ferroelectricity in marginally twisted 2D semiconductors. *Nat Nanotechnol* **17**, 390-395 (2022).

16 J. Sung et al. Broken mirror symmetry in excitonic response of reconstructed domains in twisted MoSe(2)/MoSe(2) bilayers. *Nat Nanotechnol* **15**, 750-754 (2020).

17 J. Liang et al. Resolving polarization switching pathways of sliding ferroelectricity in trilayer 3R-MoS(2). *Nat Nanotechnol* **20**, 500-506 (2025).

18 R. Bian et al. Developing fatigue-resistant ferroelectrics using interlayer sliding switching. *Science* **385**, 57-62 (2024).

19 J. Xiao et al. Berry curvature memory through electrically driven stacking transitions. *Nature Physics* **16**, 1028-1034 (2020).

20 T. H. Yang et al. Ferroelectric transistors based on shear-transformation-mediated rhombohedral-stacked molybdenum disulfide. *Nature Electronics* **7**, 29-38 (2024).

21 P. Meng et al. Sliding induced multiple polarization states in two-dimensional ferroelectrics. *Nat Commun* **13**, 7696 (2022).

22 P. He et al. Giant field-tunable nonlinear Hall effect by Lorentz skew scattering in a graphene moiré

superlattice. *arXiv:2511.03381* (accessed 2025-11-05).
23 N. J. Zhang et al. Angle-resolved transport non-reciprocity and spontaneous symmetry breaking in twisted trilayer graphene. *Nature Materials* **23**, 356-362 (2024).
24 P. He et al. Graphene moire superlattices with giant quantum nonlinearity of chiral Bloch electrons. *Nat Nanotechnol* **17**, 378-383 (2022).
25 J. Duan et al. Giant Second-Order Nonlinear Hall Effect in Twisted Bilayer Graphene. *Phys Rev Lett* **129**, 186801 (2022).
26 M. Huang et al. Giant nonlinear Hall effect in twisted bilayer WSe(2). *Natl Sci Rev* **10**, nwac232 (2023).
27 M. Huang et al. Intrinsic Nonlinear Hall Effect and Gate-Switchable Berry Curvature Sliding in Twisted Bilayer Graphene. *Physical Review Letters* **131**, 066301 (2023).
28 J. Zhong et al. Effective Manipulation of a Colossal Second-Order Transverse Response in an Electric-Field-Tunable Graphene Moiré System. *Nano Letters* **24**, 5791-5798 (2024).
29 T. Ahmed et al. Second-Order Conductivity Probes a Cascade of Singularities in a Moire Superlattice. *ACS Nano* **19**, 24930-24937 (2025).
30 S. Sinha et al. Berry curvature dipole senses topological transition in a moiré superlattice. *Nature Physics* **18**, 765-770 (2022).
31 E. Redekop et al. Direct magnetic imaging of fractional Chern insulators in twisted MoTe(2). *Nature* **635**, 584-589 (2024).
32 H. Park et al. Observation of fractionally quantized anomalous Hall effect. *Nature* **622**, 74-79 (2023).
33 N. R. Chebrolu, B. L. Chittari & J. Jung. Flat bands in twisted double bilayer graphene. *Physical Review B* **99**, 235417 (2019).
34 F. Wu et al. Giant Correlated Gap and Possible Room-Temperature Correlated States in Twisted Bilayer MoS_2. *Phys Rev Lett* **131**, 256201 (2023).
35 M. Koshino. Band structure and topological properties of twisted double bilayer graphene. *Physical Review B* **99**, 235406 (2019).
36 J. Zhu et al. Probing band topology in ABAB- and ABBA-stacked twisted double bilayer graphene. *Physical Review B* **112**, L081108 (2025).
37 P. Rickhaus et al. Correlated electron-hole state in twisted double-bilayer graphene. *Science* **373**, 1257-1260 (2021).
38 L. Liu et al. Observation of First-Order Quantum Phase Transitions and Ferromagnetism in Twisted Double Bilayer Graphene. *Physical Review X* **13**, 031015 (2023).
39 D. Waters et al. Topological flat bands in a family of multilayer graphene moire lattices. *Nat Commun* **15**, 10552 (2024).
40 T. Y. Zhao et al. Gate-Tunable Berry Curvature Dipole Polarizability in Dirac Semimetal Cd_3As_2. *Phys Rev Lett* **131**, 186302 (2023).
41 X. G. Ye et al. Control over Berry Curvature Dipole with Electric Field in WTe_2. *Phys Rev Lett* **130**, 016301 (2023).
42 I. Sodemann & L. Fu. Quantum Nonlinear Hall Effect Induced by Berry Curvature Dipole in Time-Reversal Invariant Materials. *Physical Review Letters* **115**, 216806 (2015).
43 S. Lai et al. Third-order nonlinear Hall effect induced by the Berry-connection polarizability tensor. *Nat Nanotechnol* **16**, 869-873 (2021).
44 H. Y. Liu et al. Berry connection polarizability tensor and third-order Hall effect. *Physical Review B* **105**, 045118 (2022).

45 X. Sha et al. Sign reversal of Berry curvature triple driven by magnetic phase transition in a ferromagnetic polar metal. *arXiv:2503.04616* (accessed 2025-04-06).
46 S. Sankar et al. Experimental Evidence for a Berry Curvature Quadrupole in an Antiferromagnet. *Physical Review X* **14**, 021046 (2024).
47 C.-P. Zhang, X.-J. Gao, Y.-M. Xie, H. C. Po & K. T. Law. Higher-order nonlinear anomalous Hall effects induced by Berry curvature multipoles. *Physical Review B* **107**, 115142 (2023).
48 S.-Y. Xu et al. Electrically switchable Berry curvature dipole in the monolayer topological insulator WTe2. *Nature Physics* **14**, 900-906 (2018).
49 N. Wang et al. Quantum-metric-induced nonlinear transport in a topological antiferromagnet. *Nature* **621**, 487-492 (2023).
50 A. Gao et al. Quantum metric nonlinear Hall effect in a topological antiferromagnetic heterostructure. *Science* **381**, 181-186 (2023).
51 X. Y. Liu et al. Giant Third-Order Nonlinearity Induced by the Quantum Metric Quadrupole in Few-Layer WTe_2. *Phys Rev Lett* **134**, 026305 (2025).
52 Y. Yuan et al. Interplay of Landau Quantization and Interminivalley Scatterings in a Weakly Coupled Moire Superlattice. *Nano Letters* **24**, 6722-6729 (2024).
53 L. Liu et al. Quantum oscillations in field-induced correlated insulators of a moiré superlattice. *Science Bulletin* **68**, 1127-1133 (2023).
54 L. Liu et al. Isospin competitions and valley polarized correlated insulators in twisted double bilayer graphene. *Nat Commun* **13**, 3292 (2022).
55 C. Shen et al. Correlated states in twisted double bilayer graphene. *Nature Physics* **16**, 520-525 (2020).
56 J. Orenstein et al. Topology and Symmetry of Quantum Materials via Nonlinear Optical Responses. *Annual Review of Condensed Matter Physics* **12**, 247-272 (2021).
57 J. Moser, A. Verdaguer, D. Jimenez, A. Barreiro & A. Bachtold. The environment of graphene probed by electrostatic force microscopy. *Applied Physics Letters* **92**, 123507 (2008).
58 J. Sabio et al. Electrostatic interactions between graphene layers and their environment. *Physical Review B* **77**, 195409 (2008).
59 H. Wang, Y. Wu, C. Cong, J. Shang & T. Yu. Hysteresis of electronic transport in graphene transistors. *ACS Nano* **4**, 7221-7228 (2010).
60 K. Yasuda et al. Ultrafast high-endurance memory based on sliding ferroelectrics. *Science* **385**, 53-56 (2024).
61 Z. Zheng et al. Unconventional ferroelectricity in moire heterostructures. *Nature* **588**, 71-76 (2020).
62 R. Niu et al. Giant ferroelectric polarization in a bilayer graphene heterostructure. *Nat Commun* **13**, 6241 (2022).
63 G. Maffione et al. Twist-Angle-Controlled Anomalous Gating in Bilayer Graphene/BN Heterostructures. *arXiv:2506.05548* (accessed 2025-06-05).
64 L. Chen et al. Anomalous Gate-tunable Capacitance in Graphene Moiré Heterostructures. *arXiv:2405.03976* (accessed 2025-07-01).
65 R. Niu et al. Ferroelectricity with concomitant Coulomb screening in van der Waals heterostructures. *Nat Nanotechnol* **20**, 346-352 (2025).
66 T. Zhang et al. Observation of unconventional ferroelectricity in non-moiré graphene on hexagonal boron-nitride boundaries and interfaces. *arXiv:2601.05621* (accessed 2026-04-22).
67 C. Ke, F. Liu & S. Liu. Superlubric Motion of Wavelike Domain Walls in Sliding Ferroelectrics.

*Phys Rev Lett* **135**, 046201 (2025).
68 D. Xiao, W. Yao & Q. Niu. Valley-contrasting physics in graphene: magnetic moment and topological transport. *Phys Rev Lett* **99**, 236809 (2007).
69 P. Makushko et al. A tunable room-temperature nonlinear Hall effect in elemental bismuth thin films. *Nature Electronics* **7**, 207-215 (2024).
70 H. Isobe, S. Y. Xu & L. Fu. High-frequency rectification via chiral Bloch electrons. *Sci Adv* **6**, eaay2497 (2020).

**Supporting Information:**
**Ferroelectric-tunable quantum nonlinearity of chiral Bloch electrons in a moiré system**

Zitian Pan[1,2†], Jundong Zhu[1,2†], Yu Hong[1,2], Jingwei Dong[1,2], Dongxia Shi[1,2], Kenji Watanabe[3], Takashi Taniguchi[4], Luojun Du[1,2], Wei Yang[1,2*], Guangyu Zhang[1,2*]

*[1]Beijing National Laboratory for Condensed Matter Physics and Institute of Physics, Chinese Academy of Sciences, Beijing 100190, China*
*[2]School of Physical Sciences, University of Chinese Academy of Sciences, Beijing 100190, China*
*[3]Research Center for Functional Materials, National Institute for Materials Science, 1-1 Namiki, Tsukuba 305-0044, Japan*
*[4]International Center for Materials Nanoarchitectonics, National Institute for Materials Science, 1-1 Namiki, Tsukuba 305-0044, Japan*
*[†]These authors contributed equally to this work.*
**Corresponding author: wei.yang@iphy.ac.cn; gyzhang@iphy.ac.cn*

## Device fabrication

The dual-gate twisted double bilayer graphene (TDBG) device are prepared by standard "cut and stack" method[1]. Bilayer graphene and thin BN flakes are mechanically exfoliated from bulk crystals onto silicon substrates. A thin BN flake is picked up by a stamp made of PBC film placed on PDMS, and then it is used to pick up a pre-cut bilayer graphene (BLG). Then the second half of the BLG sheet is rotated by $(60 + \theta)$ degree, and picked up by the BN/BLG heterostructure. This procedure is expected to yield an ABBA-stacked TDBG. Finally, another thin BN flake and a sheet of few-layer graphene acting as bottom gate are picked up sequentially. The final stack is released onto a silicon substrate with marks and fabricated into a Hall device via electron beam lithography (EBL), reactive ion etching (RIE) with $CHF_3/O_2$ as reactive gas, and electron beam evaporation. Ti(3nm)/Au(30nm) and Cr(3nm)/Au(30nm) are used as top gate and contact electrode, respectively[2].

## Electrical measurement

The device is wire-bonded onto a chip carrier and placed into a helium-4 vapor flow cryostats with a base temperature of ~1.7K and a maximum magnetic field of 9T. The measurement is carried out at 1.7K using lock-in amplifiers (Stanford Research SR830) with a frequency of 17.777Hz. The phase of lock-in amplifiers was locked at 90 degrees for the measurement of second-order nonlinear response, and 0 degrees for linear measurement. The real part is at least one order of magnitude larger than the imaginary part. The gate voltage is supply by two source-meters (Keithley 2400) for static measurement.

The real-time measurement of voltage pulse and device response is conducted by an oscilloscope (Keysight DSO-X 3054A), as shown in Fig. S1. Input resistance of the oscilloscope is 1MΩ. Voltage pulse is supply by a source measure unit (Yokogawa

GS610) and DC current is provided by a source-meter.

**High quality of the device**

Landau fan diagram obtained from different Hall bars all show clear and sharp quantum oscillation (Fig. S2), and the extracted twisted angle are nearly the same (Table S1), suggesting high quality and uniformity of the device.

Fig. S3 shows color maps of longitudinal resistance $R_{xx}$ as a function of carrier density $n$ and magnetic field $B$ under finite displacement field $D$. The degeneracy of landau levels out from $\nu = 0$ under low magnet field is 4, consistent with previous report[3], and is completely eliminated under high magnetic field.

Fig. S4 shows color maps of $R_{xx}$ as a function of $n$ and $D$ under $B = 0\,\mathrm{T}$ in opposite scanning direction. A sudden shift corresponds to a constant $V_t$, instead of $D$, is observed.

Fig. S5 shows color maps of $R_{xx}$ and $R_{xy}$ as a function of $n$ and $D$ under $B = 2\,\mathrm{T}$. Clear oscillations are observed, which are originated from the interplay of Landau quantization and interminivalley scatterings, suggesting the high quality of the device[3]. The sign reversal of $R_{xy}$ near $n = \pm 5 \times 10^{12}\mathrm{cm}^{-2}$ corresponds to the Van Hoff singularity.

| Area | 1 | 2 | 3 | 4 |
|---|---|---|---|---|
| Twisted Angle | 2.227° | 2.226° | 2.230° | 2.228° |

**Table S1.** Twisted angle of different area.

**Extended data for thermodynamics of the ferroelectricity**

Fig. S6 shows critical width $t_c$ to switch the polarization as a function of the high voltage level of the pulse $V_H$ under different temperature. The experimental data can be well fitted by the equation $t_c = t_0 \exp\left(-\frac{a}{V_H - V_0}\right)$ in all temperature condition.

Fig.S7 displays transfer curves of $V_t$ in opposite scanning under different temperature. The hysteresis and extracted polarization remain unchanged across 1.7-200K.

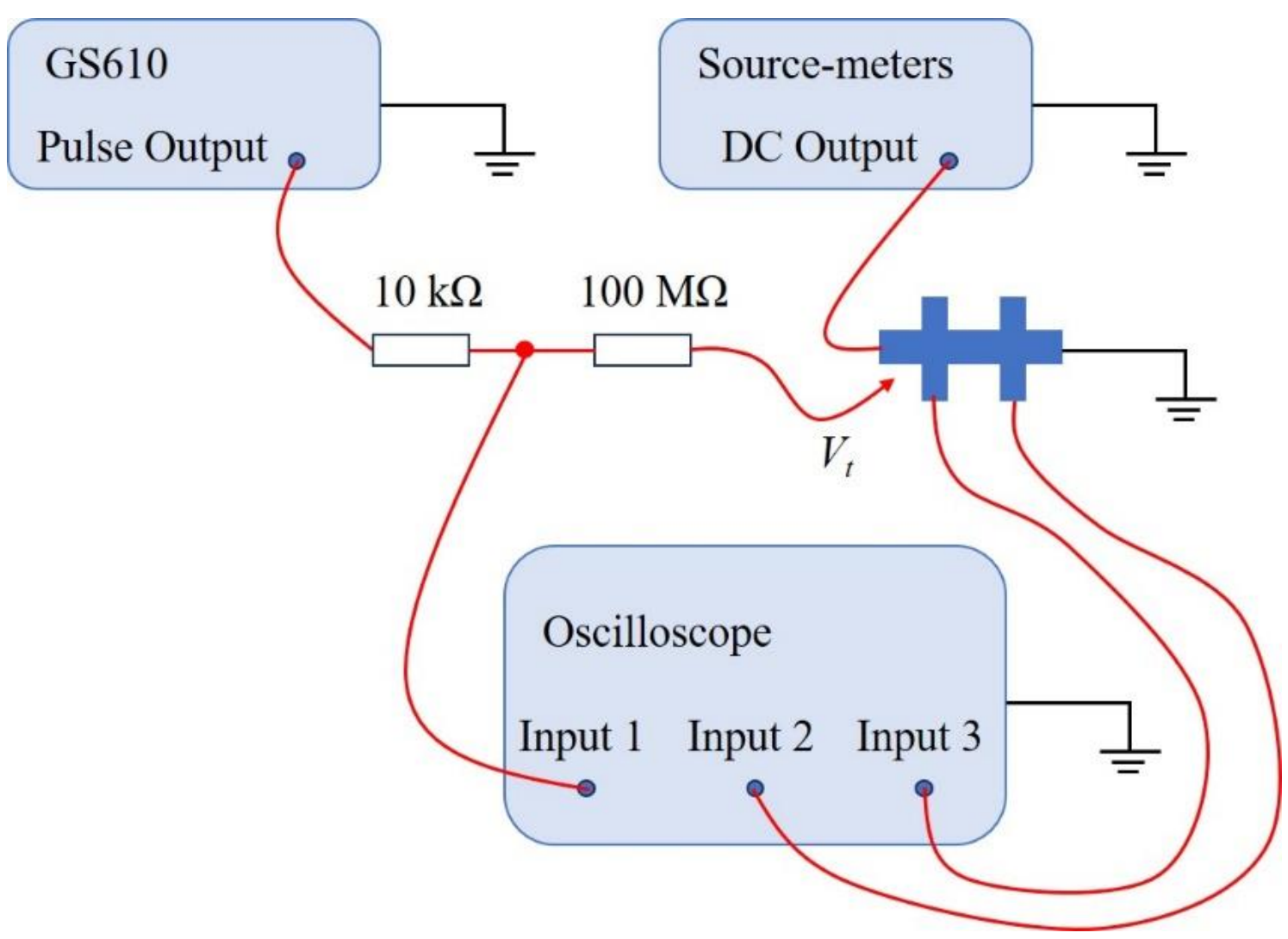

**Fig. S1** Configuration for real-time measurement of voltage pulse and device response.

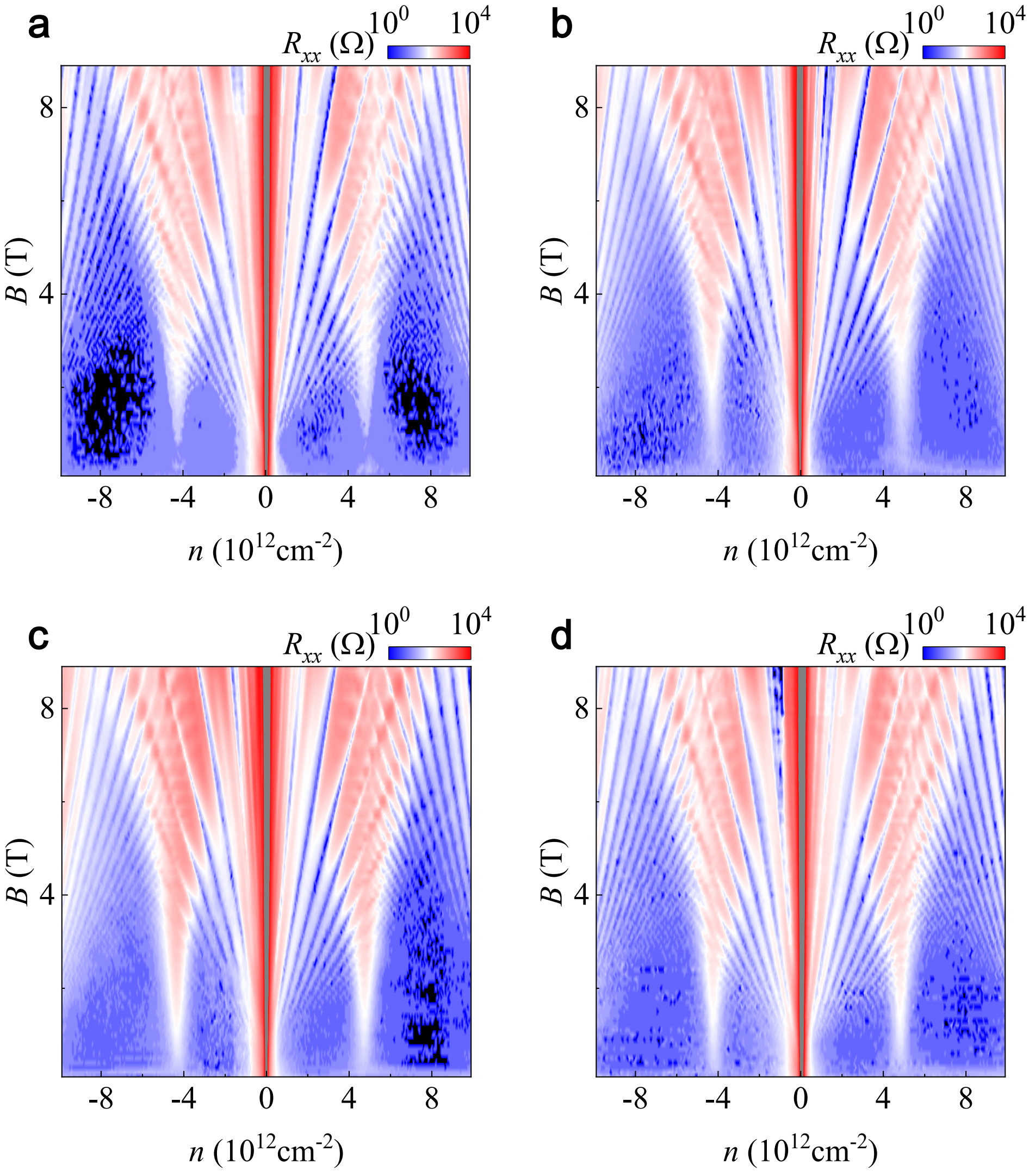

**Fig. S2 a-d,** Landau fan diagrams at $D = 0$ obtained from different Hall bars.

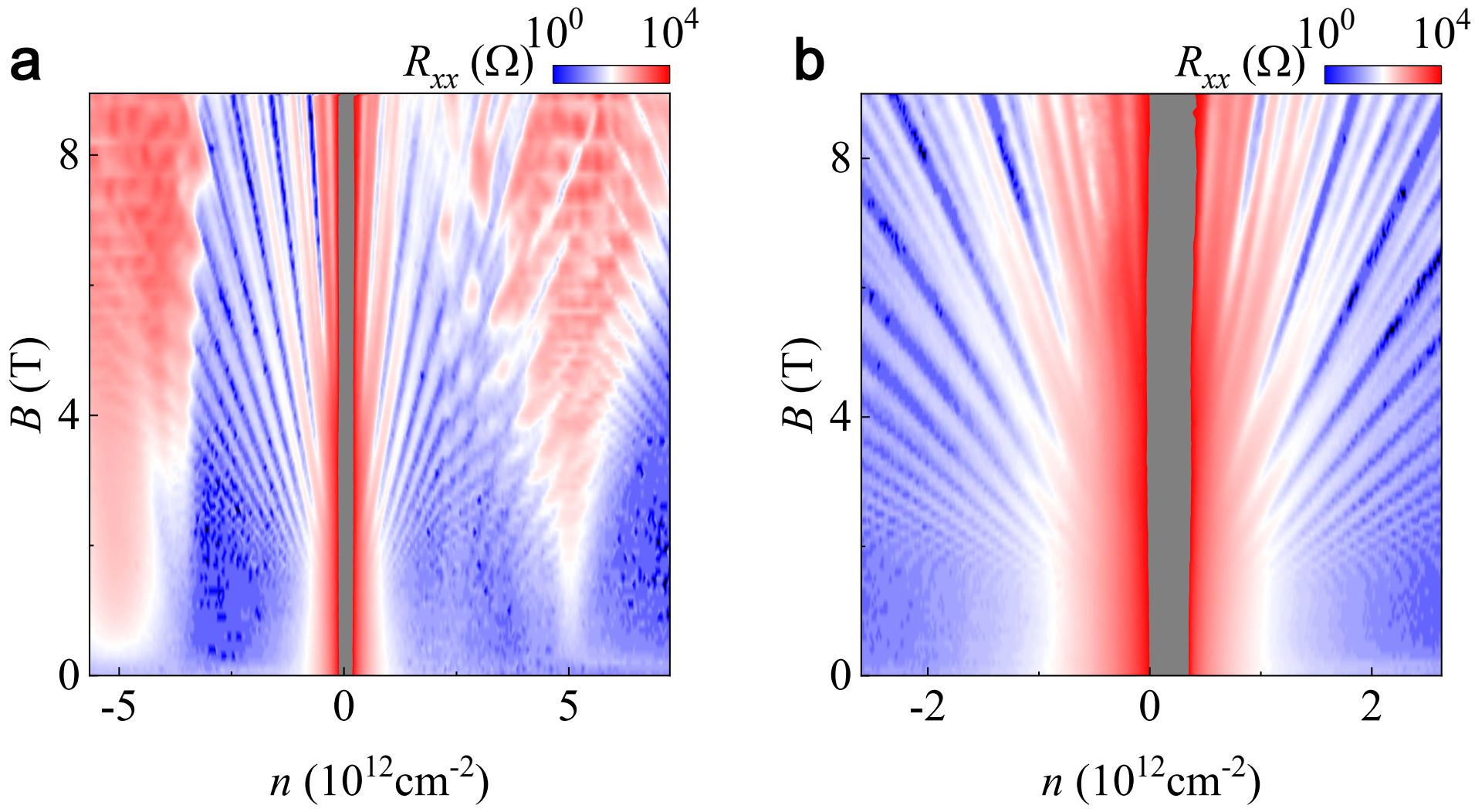

**Fig. S3 a,b,** Color maps of $R_{xx}$ as a $n$ and $B$ under displacement field of $D = -0.5$ V/nm (**a**) and $D = 0.8$ V/nm (**b**).

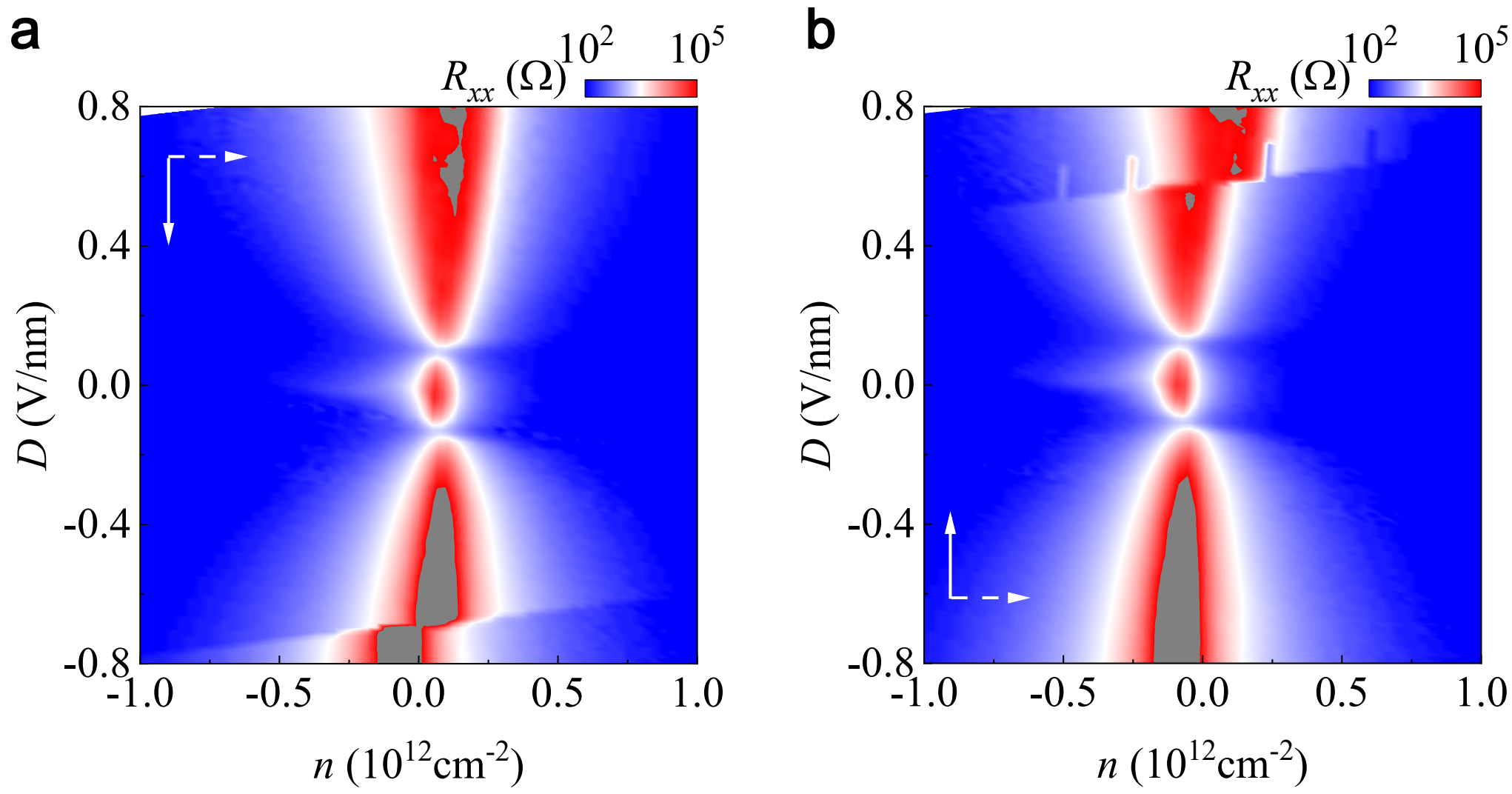

**Fig. S4 a,b,** $R_{xx}$ maps as a function of $n$ and $D$ in opposite scanning direction. The arrows indicate the scanning direction, where the solid one is fast scan and the dashed one is slow scan.

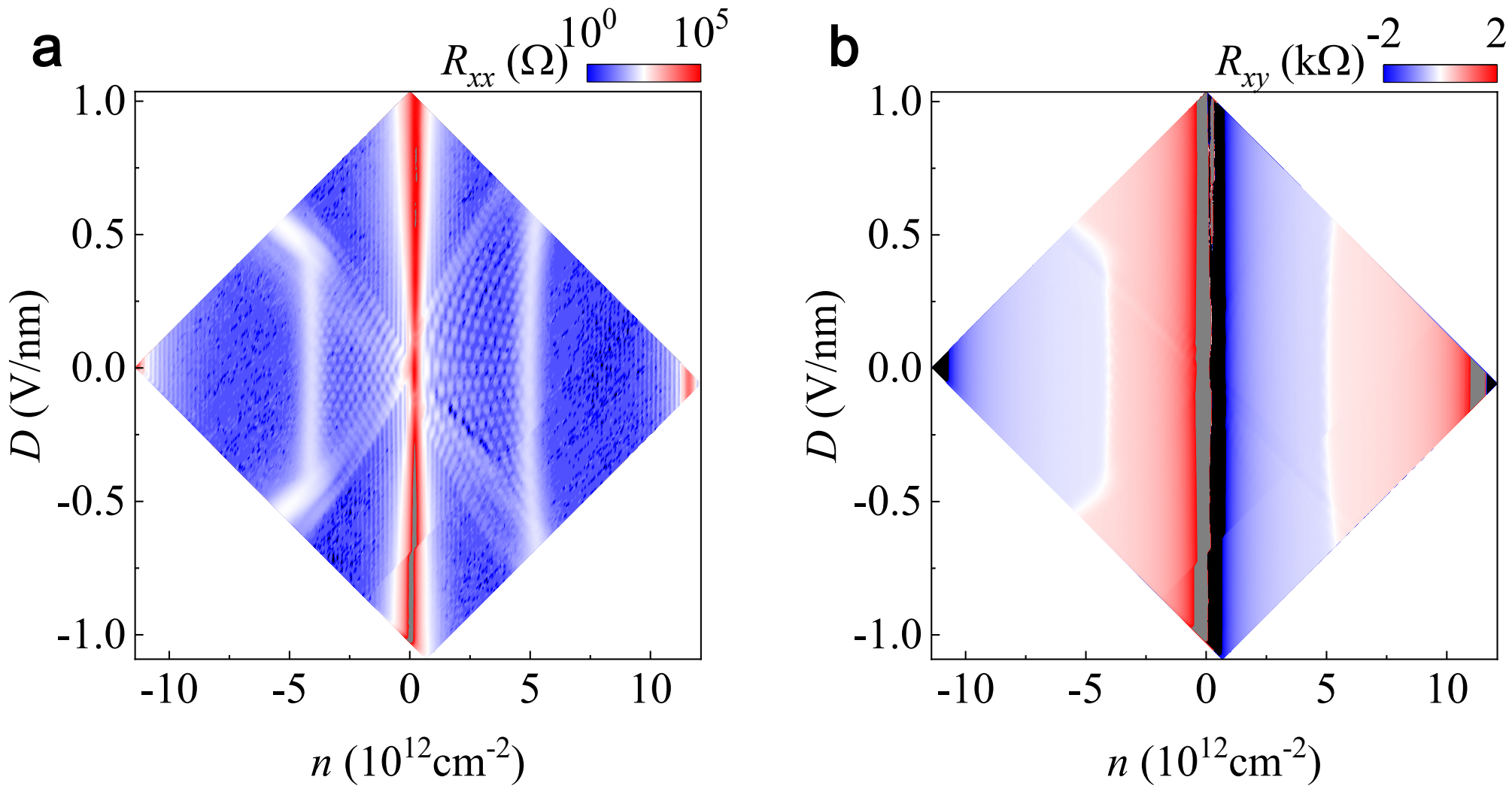

**Fig. S5 a,b,** Color maps of longitudinal resistance $R_{xx}$ (**a**) and $R_{xy}$ (**b**) as a function of carrier density *n* and displacement field $D$ under magnetic field of $B = 2$ T.

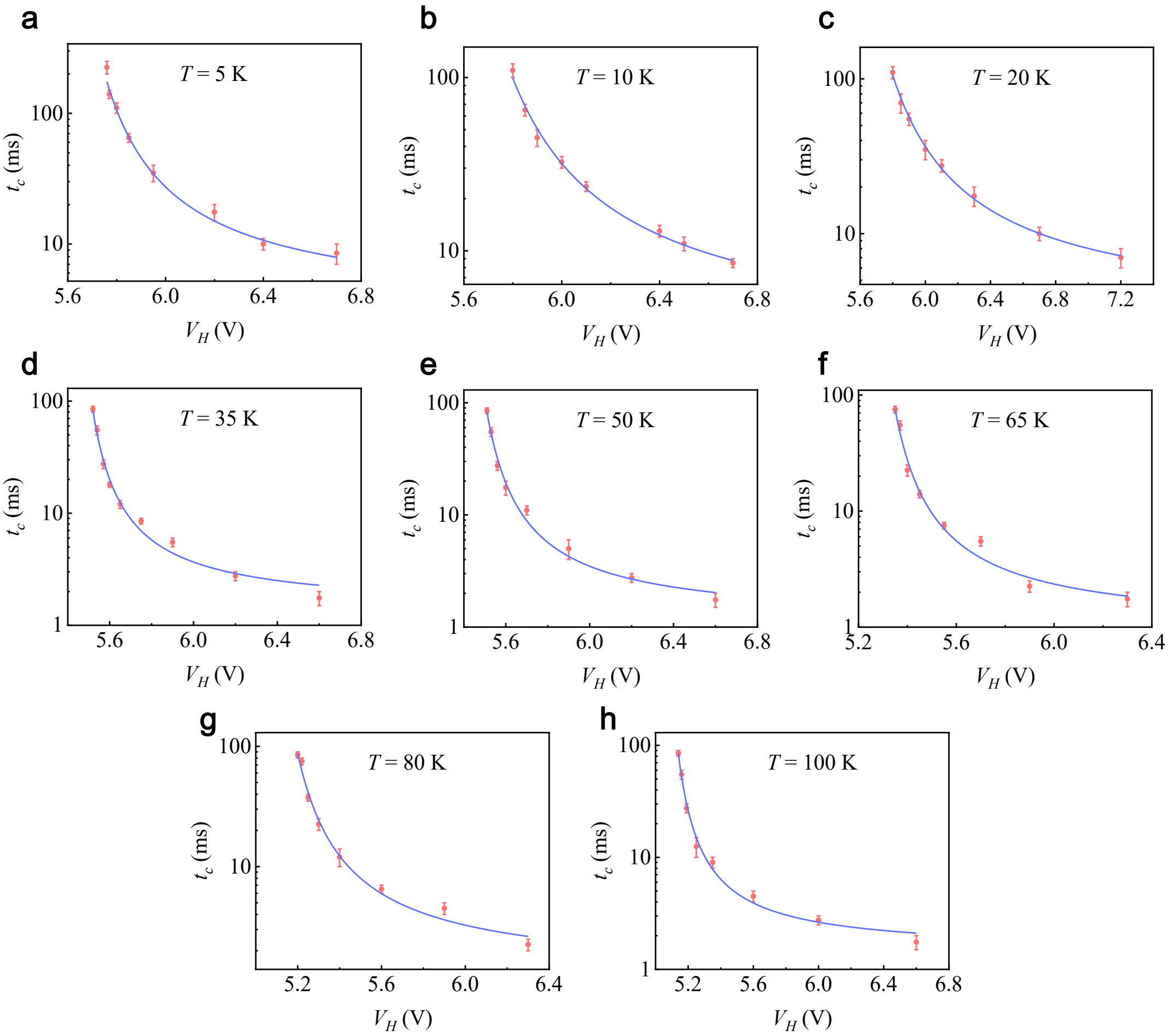

**Fig. S6 a-h,** Critical width $t_c$ to switch the polarization as a function of the high voltage level of the pulse $V_{th}$ under different temperature. The blue line is the fitting result with $t_c = t_0 \exp\left(-\frac{a}{V_H - V_0}\right)$.

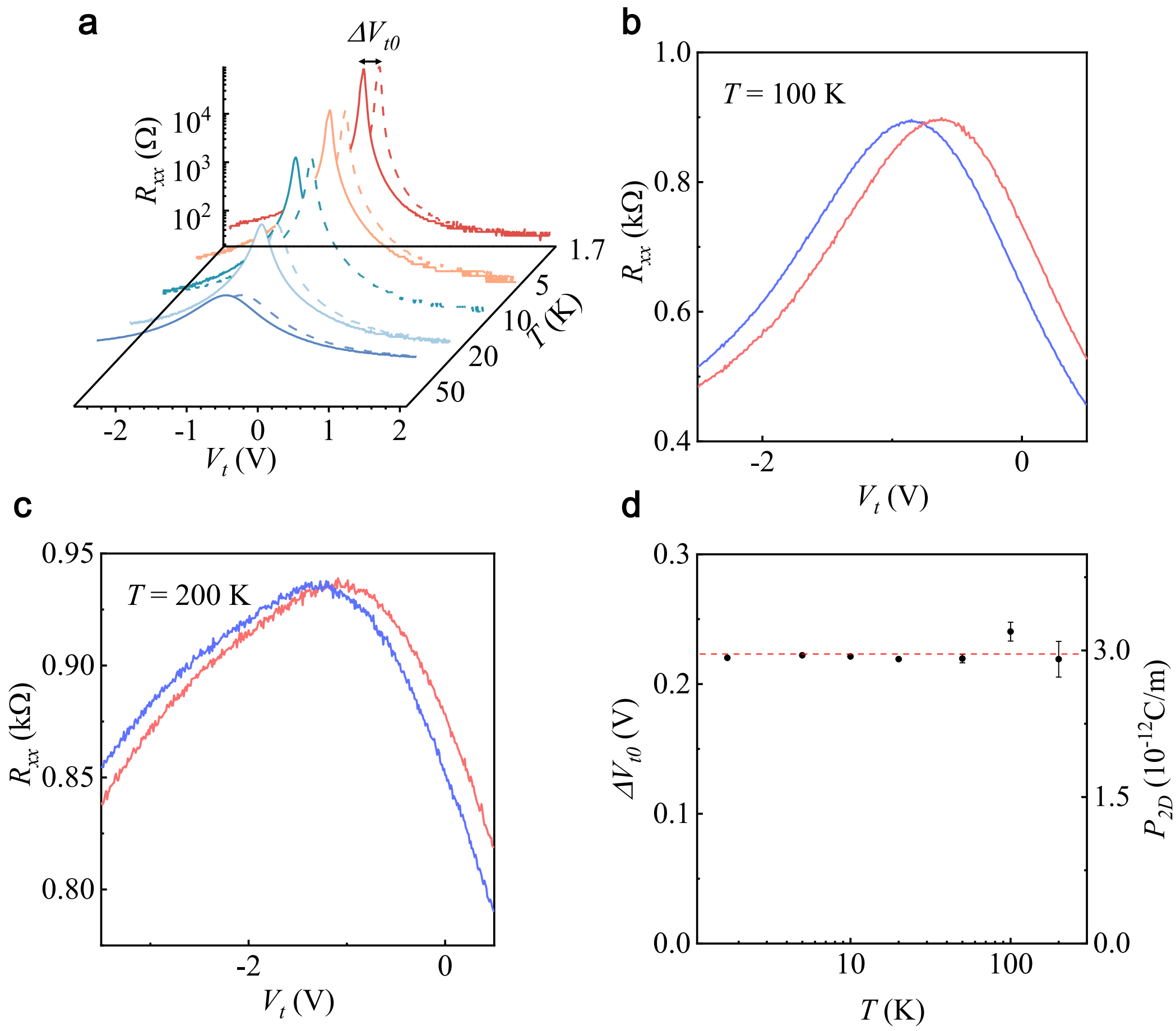

**Fig. S7 a-c,** Hysteresis of $V_t$ under different temperature. **d,** Extracted $\Delta V_{t0}$ and polarization $P_{2D}$ versus $T$.

1 K. Kim et al. van der Waals Heterostructures with High Accuracy Rotational Alignment. *Nano Lett* **16**, 1989-95 (2016).

2 L. Wang et al. One-dimensional electrical contact to a two-dimensional material. *Science* **342**, 614-7 (2013).

3 Y. Yuan et al. Interplay of Landau Quantization and Interminivalley Scatterings in a Weakly Coupled Moire Superlattice. *Nano Lett* **24**, 6722-6729 (2024).